\documentclass[11pt]{article}
\usepackage{geometry}                
\usepackage[parfill]{parskip}    
\usepackage{graphicx}
\usepackage{amssymb}
\usepackage{amsmath}
\usepackage{multirow}
\usepackage[square, comma, sort&compress, numbers]{natbib}

\title{Hierarchy of Gene Expression Data is Predictive of Future
Breast Cancer Outcome}
\author{Man Chen and Michael W. Deem\\
Department of Physics \& Astronomy\\
Rice University\\
Houston, TX  77005}

\begin{document}
\maketitle

\begin{abstract}

We calculate measures of hierarchy in gene and tissue networks of breast
cancer patients. We find that 
the likelihood of metastasis in the future
is correlated with increased values of 
network hierarchy
for expression networks of cancer-associated genes,
due to correlated expression of cancer-specific pathways.
Conversely, 
future metastasis and quick relapse times are negatively correlated with
values of network hierarchy
in the expression network of all genes, 
due to dedifferentiation of gene pathways and circuits. These results 
suggest that hierarchy of gene expression may be useful as an
additional biomarker for breast cancer prognosis.

\end{abstract}

\section{Introduction}

As cancer develops, there are changes in patterns of gene expression.  There are several examples where a defect in a single gene causes a genetic predisposition to developing cancer, for example the BRCA1 and BRCA2 genes in breast cancer \cite{Futreal1994, Miki1994, Wooster1995}. In general,  however, the development of cancer is the result of correlated networks of gene expression networks gone awry. That is, cancer is a systemic disease, and changes in fidelity of gene expression are signatures of cancer.  In some cases, changes in gene expression networks can determine disease outcome \cite{Taylor2009,Chuang2007,Pavlidis2002,Doniger2003,Draghici2003,Subramanian2005,Tian2005,Wei2007,Rapaport2007}.  Thus, structural features of gene expression networks may be `biomarkers' that can predict the probability of a patient developing or surviving cancer.

We here focus on the relation between metastasis and the structure of networks relevant to cancer.  Metastasis is the leading cause of cancer mortality \cite{Mehlen2006}. Once metastasis has occurred, the chance of patient survival drops dramatically \cite{DeMatteo2000}.  Clinicians use prognostic factors such as age or size of tumor at the time of tumor removal to predict the risk of recurrence \cite{DeMatteo2000}. Here, we present an analysis of the relation between breast cancer prognosis and hierarchical structure in correlations of cancer gene expression networks.  We will show that these measures of hierarchy in correlations of gene expression distinguish between non-metastatic and metastatic patient populations.  We will also show that these measures of hierarchy in gene expression are predictive of average time of relapse in breast cancer patients.

We are motivated to study hierarchy of gene expression by theory that relates hierarchy  to environmental stress and variability  \cite{Sun2007,Lorenz2011,Deem2013}.   This theory shows that when a system is  placed in a more variable environment, it will tend to become more hierarchical, if it has the ability to do so.  This occurs because hierarchy will tend to increase the adaptability of the system.   This theory predicts that expression networks of cancer-associated genes may be more hierarchical in more aggressive tumors or during metastasis due to increased correlations in cancer-associated gene pathways.  Conversely, measures of hierarchy in the network of all genes will likely decrease for more aggressive tumors or during metastasis, since cancer progression is a dedifferentiation of the entire gene network.

Measures of modularity have been defined for cancer gene and protein interaction networks. Carro \emph{et al.}\ identified transcriptional modules in a context-specific regulatory network  that controls expression of the mesenchymal signature associated with metastatic outcome \cite{Carro2010,Chuang2007}. This result identified a small regulatory module that was part of the mechanism that controlled an important phenotypic state in cancer cells. Chuang \emph{et al.}\ extracted subnetworks from protein interaction databases and found subnetworks that were significantly enriched with cancer susceptibility genes \cite{Chuang2007}. Comparison of normal and colon cancer gene networks identified changes in network structure. Oslund \emph{et al.}\ have ranked cancer genes candidates by local network structures, such as neighbor annotation \cite{Ostlund2010}. Yu \emph{et al.}\ have used signature analysis to identify multiple breast cancer modules \cite{Yu2006}. Taylor \emph{et al.}\ used co-expression of hub proteins and their partners to identify whether interactions are context specific, \emph{i.e.}\ interacting proteins are not always co-expressed, or constitutive, \emph{i.e.}\ interacting proteins are always co-expressed  \cite{Taylor2009,Chuang2007}. They found that during tumor progression, hub proteins are disorganized by loss of coordinated co-expression of components.  Thus, changes in the correlation of tumor interactomes were shown to be a prognostic signature in cancer.  Other studies have also demonstrated that modularity in the protein-protein interaction network or cell-cell interaction network is an important indicator for cancer prognosis \cite{Taylor2009} or tumor metastasis \cite{Balkwill2004}.

We here quantify the hierarchical structure in cancer networks, generalizing the concept of modularity. Modularity is one measure of the structure of cancer networks. Hierarchy is a measure of the modularity that exists in cancer networks at different scales.  The rest of the paper is organized as follows. In section 2 we describe how we created gene and tissue networks from gene expression data previously collected from a population of metastatic and non-metastatic patients.  In section 3, we show that hierarchy in networks of cancer-associated genes is positively correlated with metastasis due to activation of cancer specific pathways. Conversely,  for networks of all genes, we show that a measure of hierarchy is negatively correlated with metastasis and early recurrence times due to dedifferentiation. We discuss these results in section 4. We conclude in section 5.

\section{Methods}

We used gene expression profiles of breast cancer patients to construct the networks. The expression profiles were previously obtained from 286 women with lymph node negative disease who had not received adjuvant systemic treatment \cite{Wang2005,Wang2005note}. 
In that experiment, total RNA of
frozen tumor samples was hybridized to Affymetrix Human U133a GeneChips.
Expression values were calculated by the Affymetrix GeneChip analysis software
MAS 5.0.  Of the many genes analyzed, 76 cancer associated genes
were identified as predictive of metastasis.
Relapse and metastasis of the patients
were examined during follow-up visits within 5 years \cite{Wang2005}.

Of  these patients,  179 did not relapse and metastasize, and 107 were identified to have developed a distant metastasis during a follow up visit within 5 years\cite{Wang2005}.
 We seek to distinguish using the
data at year 0, the 179 patients that were disease free after extended treatment from the 107 patients that developed distant metastases within 5 years. We constructed cancer networks with two types of nodes: cancer-associated genes or tissue types.

\subsection{Gene Networks}

To construct networks of the first type, we defined a network with nodes of cancer-associated genes.   A total of 76 genes were previously identified as markers that discriminated patients who developed distant metastases from those remaining metastasis-free for 5 years \cite{Wang2005}.  We use these 76 cancer-associated genes as the nodes of our network.  The links between pairs of nodes were defined by the Pearson correlation coefficient of the two gene expression values:
\begin{equation}
l_{\alpha,\beta}=\sum_i^n\frac{(P_{i,\alpha}-\mu_{\alpha})(P_{i,\beta}-\mu_{\beta})}{\sigma_{\alpha}\sigma_{\beta}}
\label{1}
\end{equation} 
From this definition $l_{\alpha,\beta}$ is symmetric in $\alpha$ and $\beta$, and so the
graph is undirected.
Here $P_{i,\alpha}$ is the expression data of gene $\alpha$ for patient $i$ from \cite{Wang2005}, $\mu_{\alpha}$ is the average expression value for gene $\alpha$ for the $n$ patients, and $\sigma_{\alpha}$ is the standard deviation of expression value of gene $\alpha$ for the $n$ patients. 
The expression value, $P_{i,\alpha}$, is a measure of the abundance of the transcript
reported by Affymetrix GeneChip analysis software MAS 5.0, scaled to a standard target intensity
\cite{Wang2005}.
To make comparisons between the non-metastatic and the metastatic groups, which contain different number of patients,
we randomly chose  40 patients each from the non-metastatic group and metastatic group. This random selection of patients mitigates bias due to differing patient group sizes.
We repeated the procedure 100  times, which gives us 100 networks for the metastatic group and 100 networks for the non-metastatic group. 
Error bars are calculated from this bootstrapping procedure.

\subsection{Tissue Networks}
Networks of the second type are based upon tissues.  
We used tissue expression data 
previously collected for 79 human tissues
\cite{Su2004, Su2005note}.
We are motivated to study the tissue network because during metastasis cancer spreads between and through different tissue types. 
The systemic nature of metastasis suggests gene expression in different tissue types may be informative to cancer prognosis. 
The tissue network is built with tissues as nodes and correlation of gene expression between different tissues as the link values.
Specifically, we treat each tissue as a node and built a $79 \times 79$ 
tissue network, where  the link value between tissue $i$ and tissue $j$ 
is weighted by the expression data of patient $k$, $P_{k,\alpha}$
from \cite{Wang2005} to calculate a Pearson correlation coefficient:
\begin{equation}
l_{i,j}^k=\left\langle\frac{(T_{i,\alpha} P_{k,\alpha}-\mu_{i,k}) (T_{j,\alpha} P_{k,\alpha}-\mu_{j,k})}{\sigma_{i,k}\sigma_{j,k}}\right\rangle_\alpha
\label{2}
\end{equation}
This definition is symmetric in $i$ and $j$, and so the network is undirected.
Here $\alpha$ is the gene, $T_{i,\alpha}$ is the expression level of gene $\alpha$ in tissue $i$ from \cite{Su2004},
$ \mu_{i,k}$ is the average value of $T_{i,\alpha}P_{k,\alpha} $ over all the available genes,
and $\sigma_{i,k}$ is the standard deviation of $ T_{i,\alpha}P_{k,\alpha}$.
The expression value, $T_{i,\alpha}$, is a measure of the abundance of the transcript
reported by Affymetrix GeneChip analysis software MAS 5.0, normalized by a 
global median setting
\cite{Su2004}.
 We set a cutoff for using tissue expression data, from 10\% to 90\%.
Values of $T_{i,\alpha}$ falling below the cutoff are set to zero. BioGPS was used to map the gene names from the tissue data set \cite{Su2004} and the breast cancer data set \cite{Wang2005} into NCBI IDs. 
Eq.\ (\ref{2}) is an approximation to an ideal of  a dataset
with expression data for each tissue type from each
breast cancer patient.
We will show that structure in the network defined by Eq.\ (\ref{2}) has
predictive power for probability of metastasis within 5 years.

\subsection{The $CCC$}
To quantify structure in these networks, we define a measure of the hierarchy in the networks.  Since a tree topology  is the archetypal hierarchical structure, we use a measure of hierarchy that quantifies how tree-like the network is. To calculate this measure of hierarchy, we first computed
the distance matrix defined by the network. We defined the distance between node $i$ and $j$,  $d_{ij}$,  by the square root of the commute time. The commute time is the expected time it takes a random walker to travel from one of the nodes to an other and back \cite{Saerens2004}.  The commute time between nodes $i$ and $j$ depends not only on the link value but also on all the other possible paths between nodes $i$ and $j$. Note that the commute time between two nodes of a weighted graph decreases when the number of paths connecting the two nodes increases. The commute time between two nodes also decreases when the length of any path connecting the nodes decreases. These properties make the  commute time well suited for clustering tasks. To define the commute time, we let $L$ denote the graph Laplacian, defined as $L = D - A$, 
where $A$ is the matrix of links, $A = l$ 
in Eq.\ (\ref{1}) or (\ref{2}), and the diagonal matrix
$D = {\rm diag} (A_i)$, with $A_i =\sum_j A_{ij}$.
 The commute time is obtained from $L_+$, the Moore-Penrose pseudoinverse of the graph Laplacian $L$ by  \cite{Barnett1990}
\begin{equation}
n(i, j)=V_G(e_i -e_j )^TL_+(e_i -e_j ).
\end{equation}
Here $(e_i)_j =\delta_{ij}$ and $V_G =\sum_{ij} a_{ij} $. Since  $L_+$ is symmetric and positive semidefinite,
$d_{ij} =\sqrt{n(i, j)}$ is a Euclidean distance metric, called the Euclidean commute time (ECT) distance.

\begin{figure}[tbh!]
\centering
\includegraphics[scale=0.5]{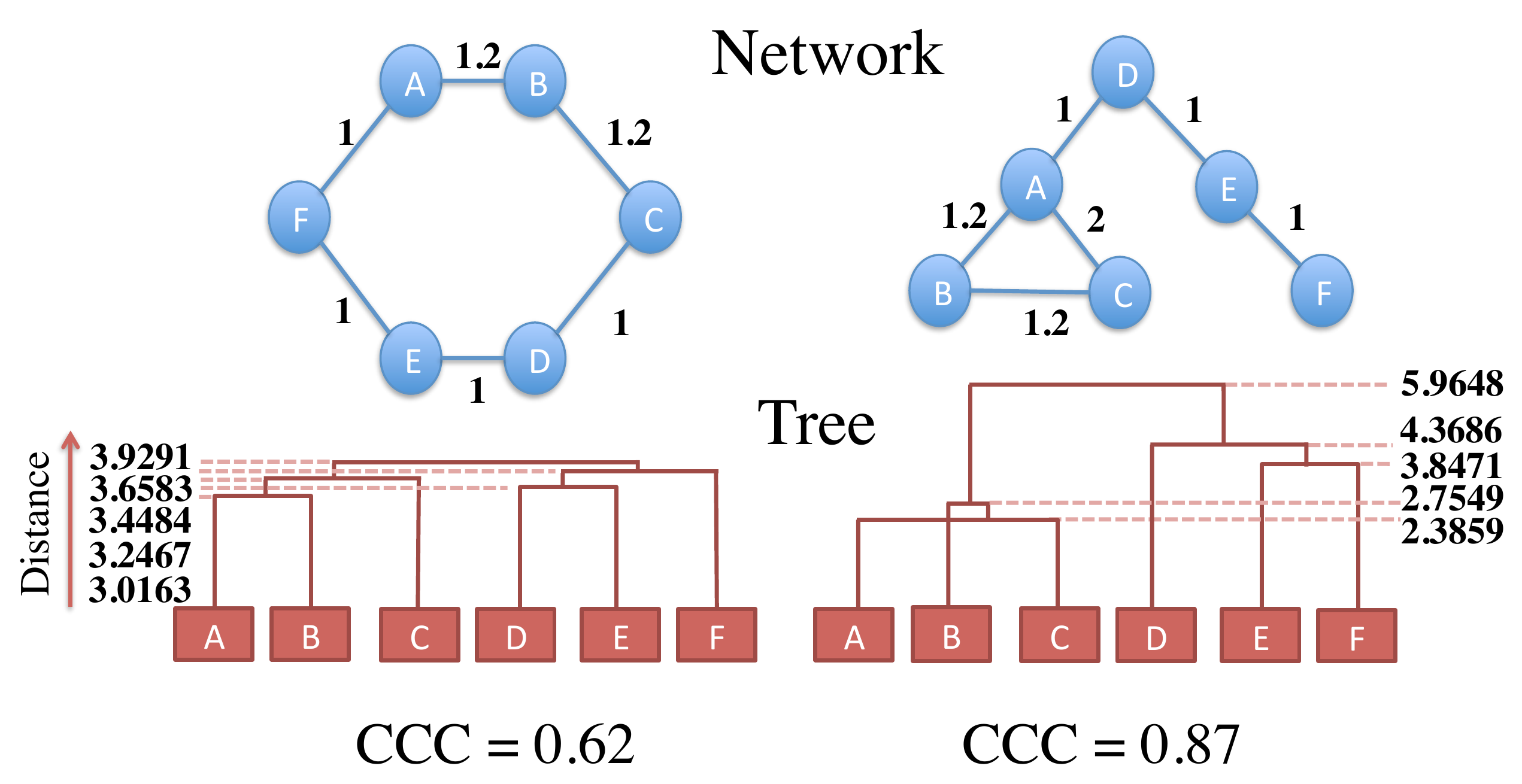}
\caption{An illustration of how the CCC is calculated, on two networks.  Distances between each pair of nodes in the network are calculated by the Euclidean commute time distance, \emph{i.e.}\ the square root of the average round trip time from one node to the other node and back. For each network, a tree best representing the network is constructed by the average linkage hierarchical clustering algorithm. 
The distance between two nodes in the tree ($y$-axis above) 
is the height above the 
baseline at which two nodes are joined in the tree topology.
To quantify the match between the tree and the original network, we calculated the correlation between the distances in the network and the distances in the tree. 
This correlation is termed the $CCC$. The more hierarchical the network, the greater the value of the CCC. 
 \label{fig1}
}
\end{figure}

We next applied the average linkage hierarchical clustering algorithm to construct the tree topology that best fit the cancer network \cite{Sokal}.
This method takes the matrix of distances between all nodes of the network, $d_{ij}$, and produces
a tree topology that best reproduces those distances.
The construction of the tree from the network by this algorithm is unique, and approximately 
optimal in reproducing the distances. 
 The distances of the tree topology are denoted by $c_{ij}$.
The tree topology has the same nodes as the original network, but different links.
We calculate the correlation between the original data and the best fitting tree, which gave the cophenetic correlation coefficient ($CCC$). The greater the correlation, the more hierarchical are the data. The nodes and links of the cancer network define the original data. The tree that best fits the data defines an approximation to the original network, termed the cophenetic matrix. The elements of the cophenetic matrix are the heights where two network nodes become members of the same cluster in the tree, see Figure \ref{fig1}.
Distance between nodes in the best fitting tree are obtained from the height of the common bifurcation point between those nodes.  This height is 
the cophenetic element of these two nodes, $c_{ij}$.  The correlation between this cophenetic matrix constructed from the best fitting tree and the original data distances is the $CCC$. The $CCC$ is a measure of similarity between the original data and the cophenetic matrix. The $CCC$ is defined as
\begin{equation}
CCC= \frac{\sum_{i<j}(d_{ij}-d)(c_{ij}-c)}{\sqrt{(\sum_{i<j}(d_{ij}-d)^2\sum_{i<j}(c_{ij}-c)^2}}
\label{ccc}
\end{equation}
Here $d$ is the average of the distances in the original network, $d_{ij}$, and $c$ is the average of the tree distances, $c_{ij}$.

\section{Results}

\subsection{Cancer-Associated Gene Network}
\begin{figure}[tbh!]
\begin{center}
a) \includegraphics[scale=0.4]{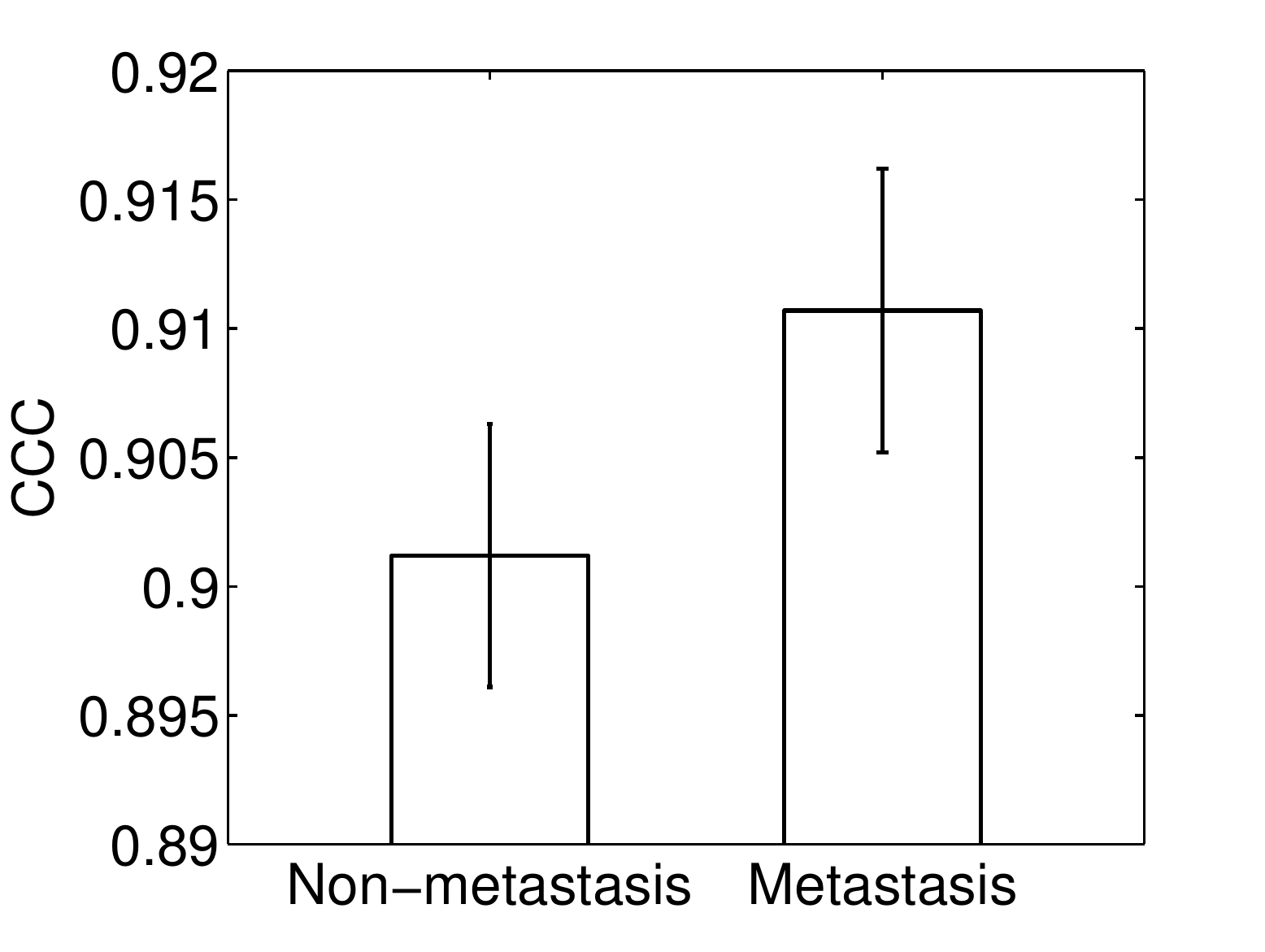}
b) \includegraphics[scale=0.4]{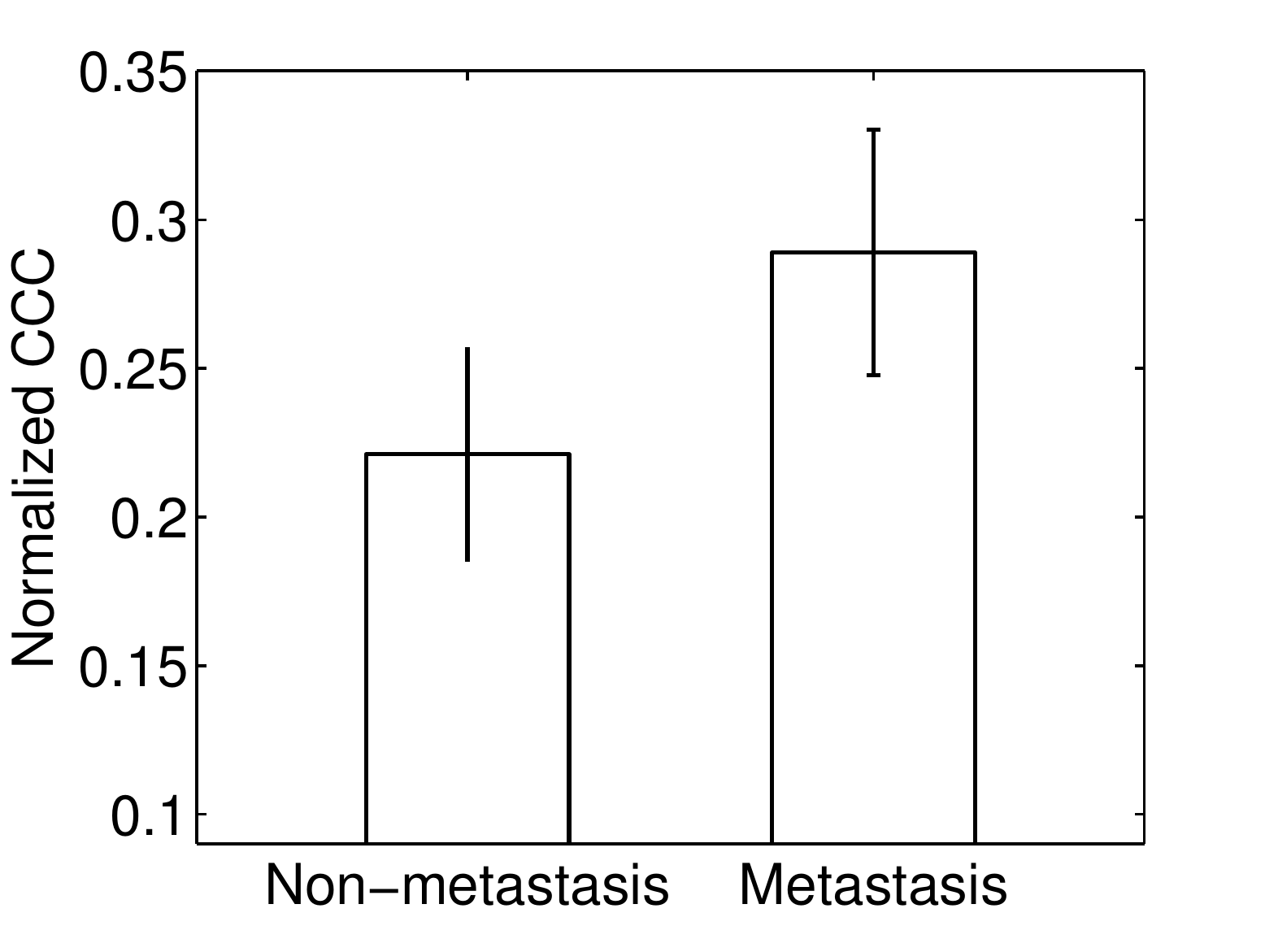}
\end{center}
\caption{a) The $CCC$ measure of hierarchy in the network of cancer-associated 
genes for the metastatic and non-metastatic patient groups.  We randomly 
choose 40 patients from each group to construct the two networks. The bootstrap
method was used to calculate the averages and standard errors,
shown by error bars.
Data are from \cite{Wang2005}.
This result shows that the network of cancer-associated genes is more 
hierarchical in the metastatic group.
b) The normalized $CCC$ in the network of cancer-associated genes for 
the metastatic and non-metastatic patient groups. 
  \label{fig2}
}
\end{figure}

For the network of cancer-associated genes, we take the 76 cancer-associated 
genes as the nodes, with link values from Eq.\ \ref{1}.  We built a network 
by constructing
a bootstrap sample of patients from either 
the non-metastatic outcome or the metastatic
outcome groups. 
We calculated the average $CCC$ value for
many 40-person networks extracted from the
bootstrap sample.
The bootstrap method was then used to calculate the overall
average $CCC$ and standard error of this average.
 Figure \ref{fig2}a shows the result:
 hierarchy of the cancer-associated gene network is greater in the metastatic group than in the non-metastatic group.

We compared these results to those from random networks. We built 100 random networks of the same size and total number of edges as the cancer-associated network, then randomly reassigned the link values in the network.
We define a normalized $CCC$ as $CCC_{\rm norm}=({CCC-CCC_{\rm rand}})/({1-CCC_{\rm rand}})$, where $CCC$ is the value of the real cancer-associated network, and $CCC_{\rm rand}$ is the average $CCC$ value of the randomized network. We computed the $z$-score of the cancer-associated network $CCC$ relative to the distribution of $CCC$ values of the random networks of the same size and sparsity,  $Z_{CCC}=({CCC-CCC_{\rm rand}})/{\sigma_{\rm rand}}$. We found $Z_{CCC}=1.64$ and  $Z_{CCC}=2.14$ for the cancer-associated gene networks in non-metastatic and metastatic patient groups compared to randomly rewired networks.

To compare the network structure between 76 cancer-associated genes with
 the network structure of the other genes, we randomly chose 76 genes from
 a total of 12926 genes and calculated $CCC$ for two groups, Eq.\ 1.
In particular, we constructed a bootstrap sample of all 12926 genes, and
then calculated the average $CCC$ for networks of 76 randomly chosen
genes from this bootstrap sample. 
 The link value is the Pearson correlation coefficient for each pair of 
76 genes for two groups. We repeated the procedure 1000 times. 
 The bootstrap method was when
used to calculate an average $CCC$ for all genes and the standard
error of this average.
The $CCC$ 
for the non-metastatic patients group of 0.925 with standard deviation 
0.0125,  compared to the $CCC$ for the metastatic patients group is 0.918 with standard deviation 0.0148. The difference for a student's t-test is significant , $p$-value $ = 1.05 \times 10^{-3}$.

\subsection{Tissue-Tissue Network}

\begin{figure}[tbh!]
\centering
\includegraphics[scale=0.4]{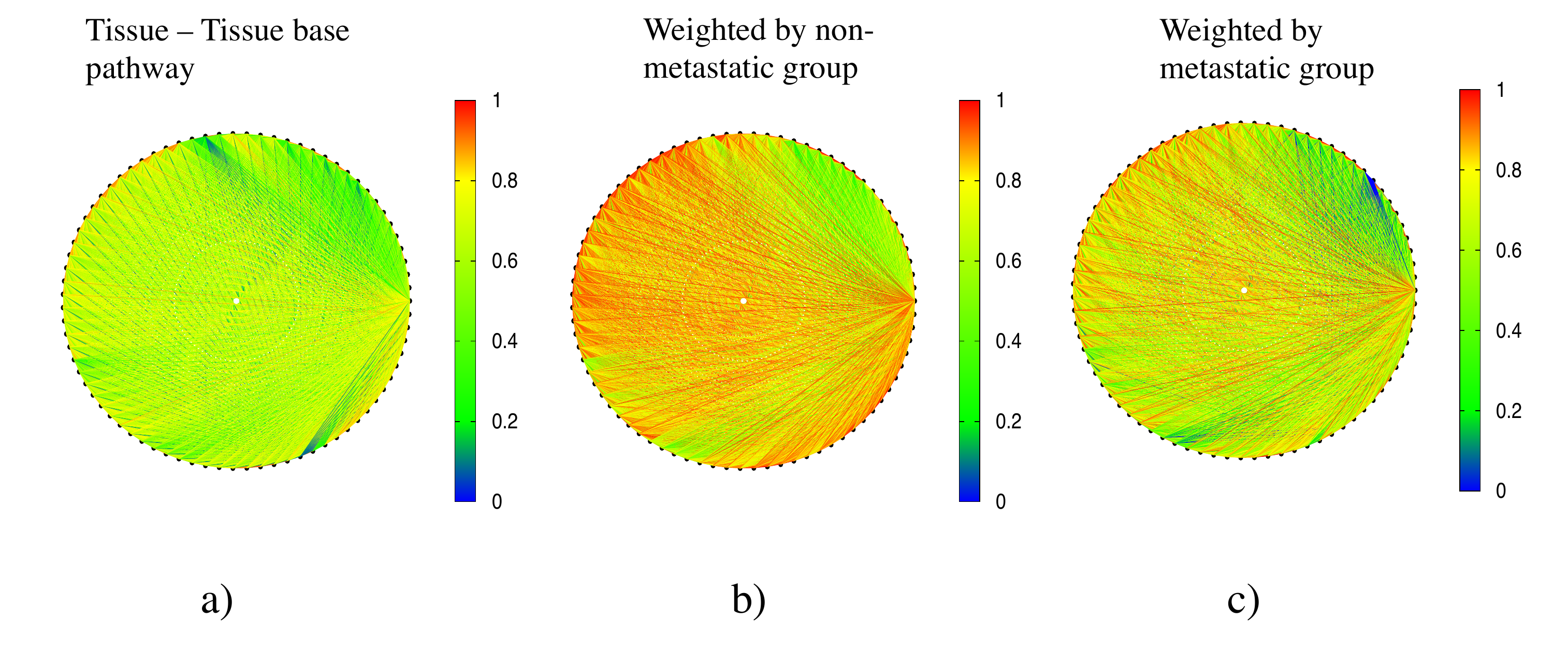}

\caption{The tissue networks with nodes as tissues and links calculated from gene expression values. a) Link value from tissue-tissue database. b) Link value weighted by the gene expression data from a patient from the non-metastatic outcome group. c) Link value weighted by the gene expression data from the metastatic outcome group. Here only those genes with the $t_{c}=0.1$ highest expression values are used.
\label{fig4}
}

\end{figure}

\begin{figure}[tbh!]
\centering
a) \includegraphics[height=2in]{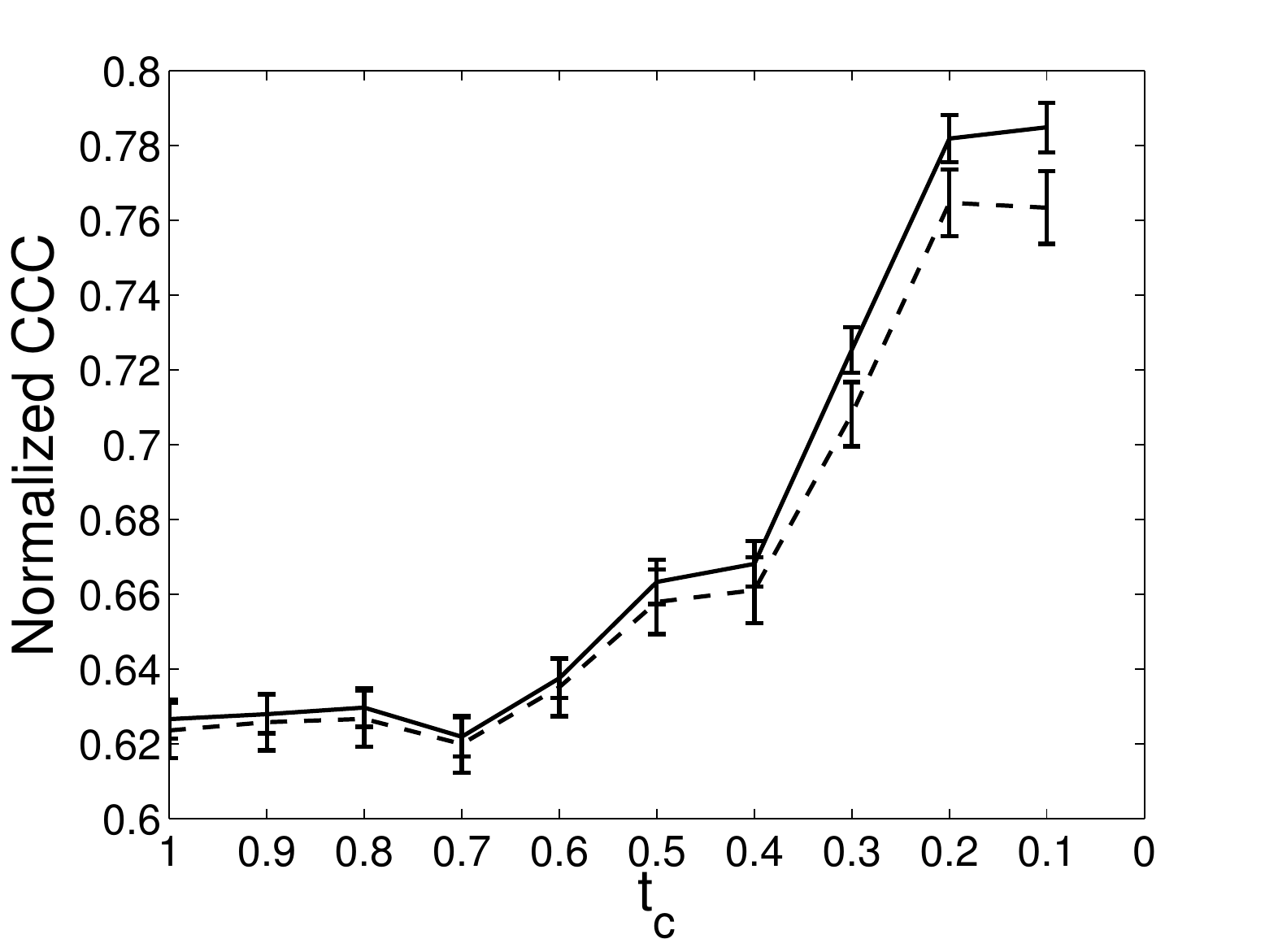}
b) \includegraphics[height=2in]{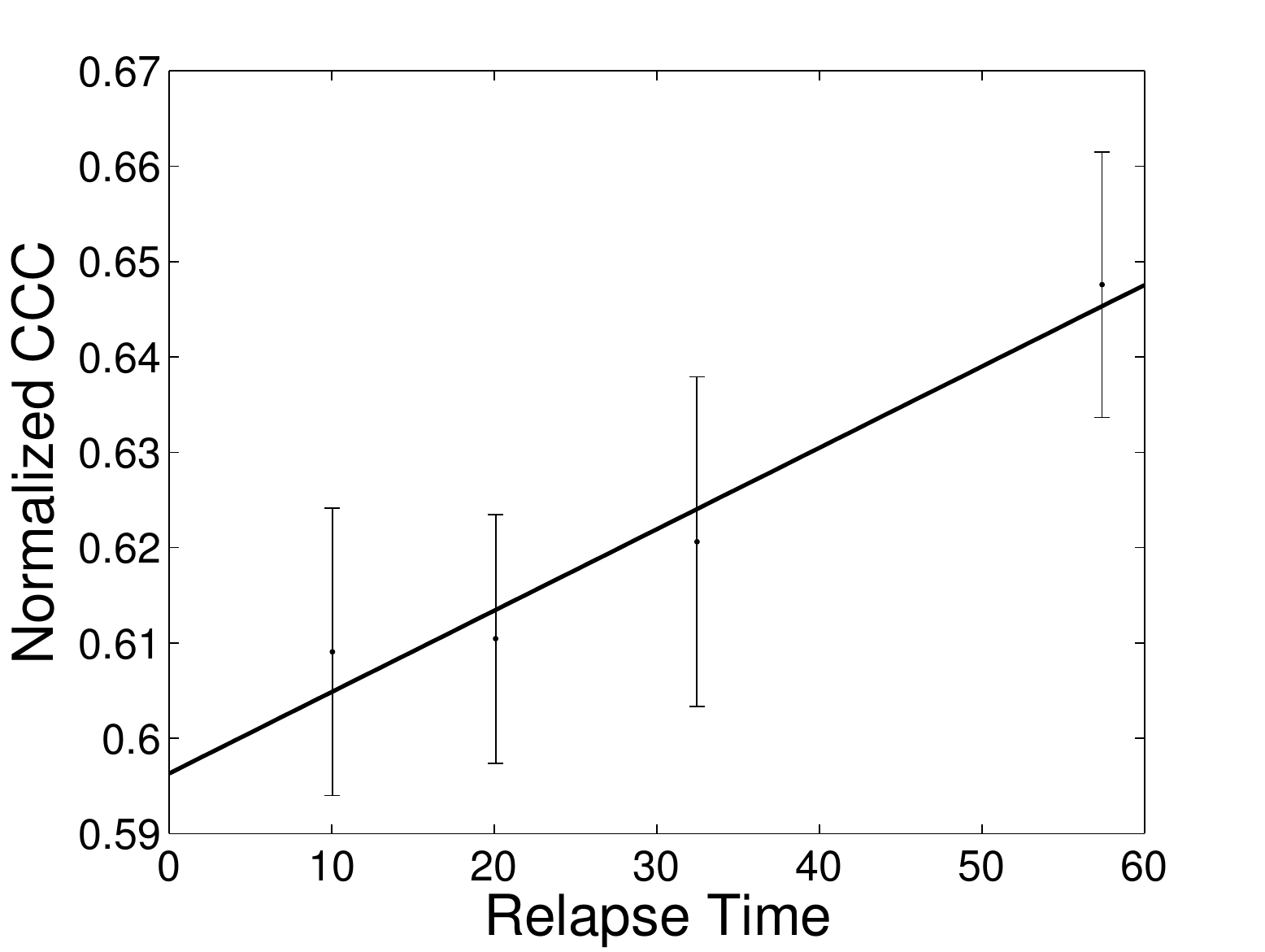}
\caption
{a) The average normalized $CCC$ calculated from
the tissue-tissue network for patients in metastatic (dashed) and non-metastatic groups (solid).  The $CCC$ for metastatic patients is below that for non-metastatic patients. The error bars are one standard error. The $p$-value for student-t test is 0.0295 for the $t_c = 0.1$ data point and less than $0.05$
for all three  $t_c = 0.3$--$0.1$ data points. The insert shows the distribution of expression levels within the tissue-tissue network. We use the highest $t_c$ fraction of these data in our analysis. b) The correlation between $CCC$ and relapse time in the metastatic patients. Cancer appears a dedifferentiation on the set of all gene values, and we here observe a correlation between shorter relapse times and lower $CCC$ values.}
\label{fig5}

\end{figure}

We built a tissue-tissue network for each patient, as a function of the expression level cutoff. 
Nodes are tissues, and link values are given by Eq.\ \ref{2}. 
Figure \ref{fig4} shows an example of this network for a patient from the non-metastatic outcome group and a patient from the metastatic outcome group.  
Figure \ref{fig4}a shows the values of the links in the tissue-tissue network, before scaling by the
expression data from the breast cancer patients, i.e.\ $l_{i,j}$ from Eq.\ (\ref{2}) with
$P_{k,\alpha} \equiv 1$.
Figure \ref{fig4}b shows the values of the links in the tissue-tissue network for a
patient in the non-metastatic group, i.e.\  $l_{i,j}^k$ from Eq.\ (\ref{2}) 
where $k$ is in the non-metastatic group.
Figure \ref{fig4}c shows the values of the links in the tissue-tissue network for a
patient in the metastatic group, i.e.\  $l_{i,j}^k$ from Eq.\ (\ref{2}) 
where $k$ is in the metastatic group.
Figure \ref{fig5}a shows the amount of hierarchy in the tissue-tissue network with the expression level cutoff ranging from 0.1 to 0.9 for the metastatic and non-metastatic patient groups. For each patient, we determined the time of cancer recurrence.  In typical cancer analysis, recurrence within 5 years of surgery indicates `non-cure.'  More rapid relapse times are interpreted as more aggressive cancer recurrence. Figure \ref{fig5}b shows the relationship between $CCC$ and relapse time.

The structure of the tissue-tissue network is different from that of a random network.  As with the gene network, we compared the real tissue network to a randomized network, in which link values are randomly reassigned.  The 
average $CCC$ of the randomized network in the range $t_c=0.1$ to $0.9$ is 
 $0.864$ with standard deviation 0.017.  The corresponding average value 
of the real network from which Fig \ref{fig5}a is derived
is $0.955$ with standard deviation $0.011$. The $z$-score is, therefore, $Z_{CCC}= 5.35$, which indicates that there is statistically significantly more hierarchy in the real tissue network than in the randomized network

\section{Discussion}

The classical view of cancer is that it is a dedifferentiation of the host.  A disruption of the structure of the tissue-tissue network is, indeed, observed in Figure \ref{fig5}.  The structure of the network in patients with more aggressive, metastatic cancers is more disrupted than in patients with no metastasis.  Furthermore, among the metastatic patients, the structure of the network was more disrupted in the patients with the more aggressive cancers that recurred earlier, as seen in Fig.\ \ref{fig5}b. Both of these results are consistent with the picture of cancer as a general dedifferentiation of the host tissue network.  From the point of view of the host, cancer is a disruption.  Structure in the host network, which endows the host with robust functioning, is destroyed by the cancer.

\begin{figure}[tbh!]
\centering
a) \includegraphics[height=1.3in]{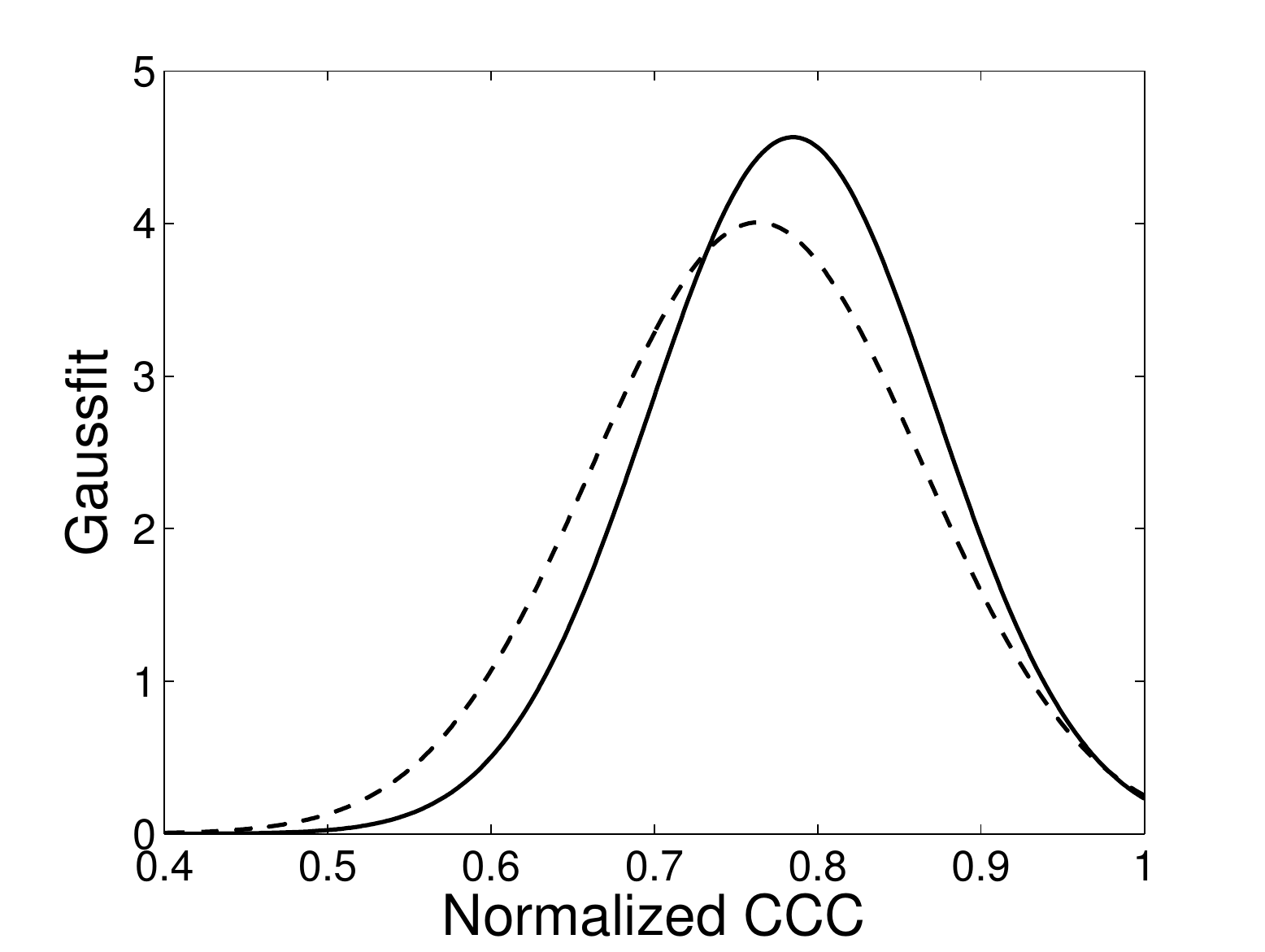}
b) \includegraphics[height=1.3in]{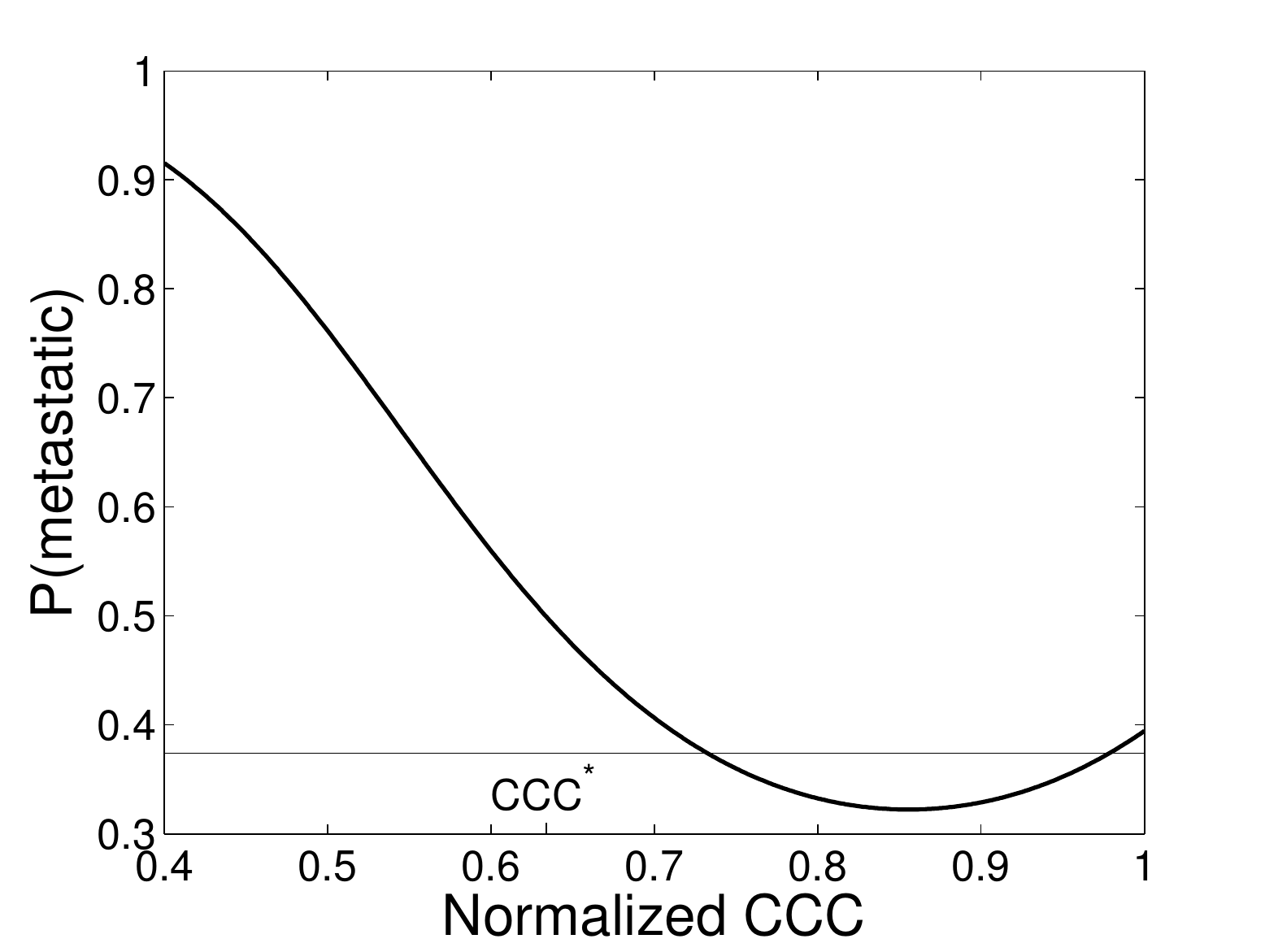}
c) \includegraphics[height=1.3in]{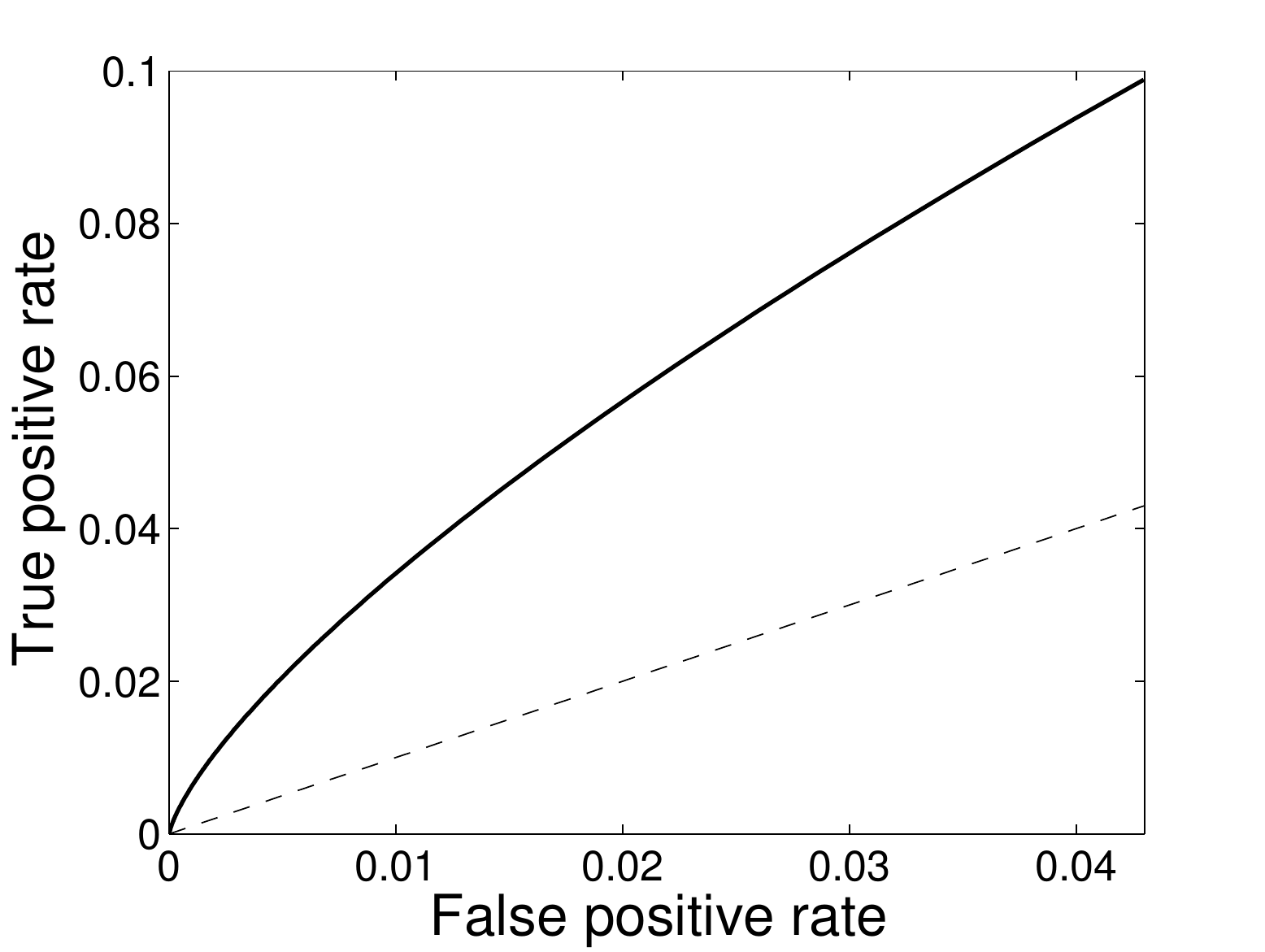}
\caption{ a) The distribution of $CCC$ values for the non-metastatic (solid) and metastatic (dashed) outcome patient populations.
We calculated the average $CCC$ and the standard derivation of the 
$CCC$ in the non-metastatic group and the metastatic groups.
We used the tissue-tissue networks with $t_c=0.1$. The resulting, Gaussian distribution fits to the non-metastatic and metastatic groups are shown. b) The probability of metastasis for
a given patient depends on the $CCC$ value for that patient, from Eq. (\ref{marker}). 
Lower values of $CCC$ are more likely to lead to metastasis.
The thin horizontal
line illustrates the 37.4\% probability of metastasis in the population, e.g.\ 107 metastatic
patients, divided by 286 total patients.
c) The ROC curve for the prediction 
that $CCC < x_{\rm cutoff}$ 
leads to metastasis,
shown only for the 
5\% of the population with the smallest $CCC$.
\label{fig6}
}
\end{figure}

Thus, we expect the values of the $CCC$ for the tissue-tissue network
to be lower for patients in the metastatic group.
Figure \ref{fig6}a shows the distribution of 
the normalized $CCC$ for each patient in the 
non-metastatic or metastatic outcome groups.  
We use $t_c = 0.1$ because for this value, there is the greatest discrimination
between the metastatic and non-metastatic populations in Figure \ref{fig5}a.
These distributions, when averaged, give the $t_c=0.1$ values in Fig \ref{fig5}a.  
The distribution of the metastatic outcome group is shifted to lower average
normalized $CCC$ values.  In addition, the width of the metastatic population distribution is slightly larger.  Figure \ref{fig6}b shows the probability of metastasis for a patient with a
 given normalized $CCC$ value, according to the equation 
\begin{equation}
p({\rm metastatic}) = \frac{N_{\rm metastatic} f_{\rm metastatic}}{
N_{\rm metastatic} f_{\rm metastatic} + N_{\rm non-metastatic} f_{\rm non-metastatic}}
\label{marker}
\end{equation} 
 Here $N_{\rm metastatic} = 107$, and $N_{\rm non-metastatic} = 179$. The values of 
$ f_{\rm metastatic}$ and $ f_{\rm non-metastatic}$ are equal to the dashed and solid curves in Fig. \ref{fig5}a, respectively.
The quantity  $p({\rm metastatic})$ is a biomarker.  The biomarker is highly discriminating for low values of the 
$CCC$, although only a small fraction of the patients have such low values of the $CCC$.  For example, 5\% of the patients values are below 
$CCC^*_{\rm norm}=0.634$, 
here
for which $p({\rm metastatic})=0.5$.

We note that the standard BRCA-1 and BRCA-2 biomarkers for cancer apply to roughly 5--10\% of women \cite{SEER, Genetics}.  Thus, the biomarker in Eq.\ \ref{marker}, which applies to only 5\% of the patients, is perhaps of more significance than it may initially appear.  Female subjects
with BRCA-1 biomarkers have a cumulative lifetime risk of breast cancer in the
range of 50--80\%, versus a background risk of 12.5\%
 \cite{SEER, Genetics}.  The predictive power in Fig.\ \ref{fig6}b, also $> 50\%$ for the 5\% of the population with $CCC < $ $CCC^*$, is, therefore, perhaps also of greater significance than one might initially think.
The $CCC$ may be combined with other genetic biomarkers to achieve
increased predictive power
 \cite{Taylor2009,Chuang2007,Pavlidis2002,Doniger2003,Draghici2003,Subramanian2005,Tian2005,Wei2007,Rapaport2007}.

In Fig.\ \ref{fig5} we find that highly expressed genes contribute more to the distinction between structures of tissue-tissue networks in patients with metastatic and non-metastatic outcomes. That is, highly expressed genes may have more impact on cancer outcome than lowly expressed genes.  We do find, for example, the average expression level of cancer-associated genes is 15.25\% higher than the average expression level of all genes.  

One view of cancer is that it is an activation of cancer-specific pathways, perhaps hijacked atavistic host pathways \cite{Davies2011}. Therefore, by examining cancer-specific pathways, we should see the development of structure in cancer patients.  Figure \ref{fig2} shows that the structure of networks of cancer-associated genes is greater in patients with aggressive, metastatic cancers than in patients with non-metastatic cancers.
Cancer is an activation on the network of cancer associated genes. 
Conversely, because cancer is a dedifferentiation on the network of all genes, 
the $CCC$ is lower for the entire gene network 
in the more aggressive tumors of
metastatic patients in Figure \ref{fig5}.
As discussed in section 3.1, the $CCC$ of the gene network for randomly chosen genes is lower in the metastatic patients than in non-metastatic patients. The results show that the network structure of cancer-associated genes correlated with the clinical outcome. That is, metastatic tumors dedifferentiate the structure of most genes, but build up the structure of cancer associated gene networks. Recalculating the tissue-tissue network, Eq.\ (\ref{2}), using only
the cancer associated genes confirms this result.
Eq.\ (\ref{2}) is used, but only for $\alpha$ within
the 6 cancer associated
genes from \cite{Wang2005} that are also present in the
dataset of \cite{Su2004}.  
As expected, this calculation shows the trend in
Figure \ref{fig5} is reversed to that of Figure \ref{fig2}: the
$CCC$ of the tissue-tissue network constructed from only cancer associated
genes is higher for the metastatic group than 
for the non-metastatic group, with $p$-value 0.076.

The $CCC$ provides a new perspective in studying the structure of cancer networks. A higher $CCC$ indicates a more hierarchical network, indicative of increased structure.  This increased structure often allows for greater evolvability and is often induced by environmental stress \cite{Sun2007}. Networks of cancer associated genes in metastatic patients are more hierarchical than in non-metastatic patients.
 
\section{Conclusion}

We have defined a measure of hierarchy in cancer networks. We found a correlation between the $CCC$ and the clinical outcome. In our study,  the $CCC$ of the cancer-associated gene network was higher for the metastatic outcome group than for the non-metastatic outcome group. 
We anticipated this result, partly because physics of evolution in changing environments \cite{Sun2007, Alon2005} suggests that increased hierarchical structure helps cancer to better adapt to the changing environments encountered in metastasis and to overcome the natural barriers to metastasis in the body. 
  
  We find highly expressed genes  play a particularly important role in predicting the metastasis of breast cancer. We found that disruption of the tissue-tissue network 
is correlated with both metastatic potential and shorter time of recurrence.
Cancer is a complex disease involving genetic, epigenetic, and environmental perturbations. 
Furthermore, cancer operates within and between tissues.
Our study of the tissue-tissue network provides additional insights and a possible additional biomarker for breast cancer metastasis and recurrence. 

\section*{Acknowledgments}
This work was supported in part by the US National Institutes of Health
under grant
 1 R01 GM 100468--01.
\bibliographystyle{unsrt}
\bibliography{cancerpaper}

\end{document}